\documentstyle[aps,prl,epsf,twocolumn,floats,amssymb,amsfonts]{revtex}

\newcommand{\bm}[1]{\mbox{\boldmath $#1$}}

\begin{document}
\draft
\title{
Density of states for dirty $d$-wave superconductors:\\
A unified and dual approach for different types of disorder
}
\author{Claudio Chamon$^{a}$ and Christopher Mudry$^{b}$}
\address{
$^{a}$Department of physics, Boston University, Boston, MA 02215, USA
\\
$^{b}$Paul Scherrer Institut, CH-5232, Villigen PSI, Switzerland\\
{\rm \today}
\medskip \\ \parbox{14cm}{\rm 
A two-parameter field theoretical representation is given of
a 2-dimensional dirty $d$-wave superconductor that interpolates between
the Gaussian limit of uncorrelated weak disorder and the unitary
limit of a dilute concentration of resonant scatterers. It is argued
that a duality holds between these two regimes from which follows
that a linearly vanishing density of states in the Gaussian limit
transforms into a diverging one in the unitary limit arbitrarily close
to the Fermi energy.
\smallskip \\
PACS numbers: 72.15.Rn, 71.23.-k, 71.55.-i}}

\maketitle 


\narrowtext

Extrinsic defects in high $T_c$ materials have observable consequences
for the local and bulk density of states (DOS) of quasi-particles.
This is seen by scanning tunnelling microscopy\cite{Hudson99},
nuclear magnetic resonance\cite{Fink90}, and
neutron scattering\cite{Fong99}.
Popular wisdom distinguishes ``weak'' extrinsic disorder 
introduced by defects above or below the copper-oxide planes
through oxygen doping, from ``strong'' extrinsic disorder caused by
substitutions of in-plane copper atoms with ions such as zinc.

Given that the symmetry of the superconducting state is
predominantly of the $d$-wave type in high $T_c$ materials, 
it is thus not surprising that a considerable theoretical effort 
has been devoted to the effect of quenched disorder on a $d$-wave 
order parameter by reducing the problem to two models.
     The first one is a model of
     Anderson localization for a single ``nodal'' quasi-particle 
     described by a 2-dimensional (2d) 
     non-magnetic random Bogoliubov-de-Gennes (BdG) Hamiltonian
\cite{Lee93,Nersesyan94,Nagaosa95,Balatsky95,Mudry96,Ziegler96,Khaliullin97,Pepin98,Senthil98,Atkinson00,Zhu00,Simons00,Huckestein00}.
     The second one is a Kondo model  
     in which a single magnetic moment
     couples to a bath of 2d ``nodal'' quasi-particles\cite{Polkovnikov00}.
At this level of approximation no attempt is made to describe either
the existence of quasi-particle interactions or
the self-consistent effect of defects on the superconducting order parameter,
although these effects might be important\cite{Atkinson00}. 
In spite of the simplicity of either models many important theoretical 
issues remain unresolved to this date.

Even in the simpler case of Anderson localization, 
the effect of disorder on the quasi-particle 
DOS of a $d$-wave superconducting state is poorly understood. 
The physics of localization is expected to dominate at 
long distances and small energies in 2d\cite{Senthil98,Simons00}.
This fact together with the existence of particle-hole symmetry below $T_c$
makes it essential to account for the level repulsion 
upon approaching the Fermi energy $\varepsilon_{\rm F}$. 
However, the physics of level repulsion
and localization are non-perturbative with respect to
standard perturbative techniques
[expansions in $(k_{\rm F}\ell)^{-1}\ll1$,
$k_{\rm F}$ the Fermi momentum and $\ell$ the mean free path
in the Born approximation]\cite{Altland97}.
The existence of four isolated nodes of 
the superconducting pairing function $\Delta_{\bf k}$
in the 2d Brillouin zone also invalidates, 
upon approaching $\varepsilon_{\rm F}$, 
expansions in powers of $(k_{\rm F}\ell)^{-1}\ll1$\cite{Nersesyan94}.
Moreover, there are some theoretical arguments and numerical evidences 
that have been attributed to the existence of nodes of $\Delta_{\bf k}$
suggesting that the dependence on energy of the DOS depends 
in a dramatic fashion on the microscopic details of the disorder.
On the one hand, for a high density of very weak $s$-wave scatterers 
[the Gaussian limit (GL)] level repulsion dominates and causes the
DOS to vanish algebraically upon approaching $\varepsilon_{\rm F}$
with an exponent that can depend on the disorder strength if the disorder
is sufficiently smooth, i.e., correlated in 
space\cite{Nersesyan94,Mudry96,Huckestein00}.
On the other hand, a dilute density of uncorrelated and 
resonant $s$-wave scatterers [the unitary limit (UL)] 
is argued to result in a diverging DOS\cite{Pepin98,Atkinson00,Zhu00}.
The purpose of this paper is to present a unified picture
of these two limits.

Starting from a BdG Hamiltonian describing 
a dirty $d$-wave superconductor with {\it white-noise correlated disorder} 
that depends on two parameters, the impurity concentration $\rho$ 
and the variance $g$ of the disorder,
we construct an effective field theory that generates 
disorder averaged Green functions.
We show how this effective action reduces to the nonlinear sigma model (NLSM)
describing a random BdG Hamiltonian with time reversal symmetry and 
spin rotation symmetry 
(class CI in the classification of Ref.~\onlinecite{Altland97})
in the limit of $\rho\to\infty$ with $g$ finite.
The effective action is treated by performing
an expansion in powers of $\rho$ (virial expansion)
in the opposite limit of $g$ fine-tuned to its resonant value
$g\to g_{\rm r}$ with $\rho$ finite.
Carrying out the expansion up to first (second) order in $\rho$
allows to identify this limit with the UL (the expansion parameter).
As was the case with the $(k_{\rm F}\ell)^{-1}$ expansion, 
the virial expansion converges only for not too small energies.
In order to better understand whether there can or cannot be a
crossover from the UL to the GL, we use a variational method and show
that the form of the self-consistent equations is controlled by how
the chemical potential is introduced in the problem. This suggests
that the energy alone cannot connect the two limits if the chemical
potential is chosen at the ``resonant'' condition for the UL. By
expanding perturbatively for {\it large} energies, we identify a
duality relation between the GL and UL series that would imply, at 
{\it low} energies, that the diverging DOS
$1/|\varepsilon|\ln^2|1/\varepsilon|$ is connected by duality to the
vanishing DOS $|\varepsilon|$ in the GL arbitrarily close to 
$\varepsilon_{\rm F}$. 
This non-perturbative argument relies in an essential way on the nodes of
$\Delta_{\bf k}$.

To model a dirty 2d $d$-wave superconductor, we use
the tight-binding Hamiltonian ${\cal H}={\cal H}_0+{\cal V}$,
\begin{eqnarray}
&&
{\cal H}_0=
-\sum_{\langle ij\rangle}
\left(
 t^{\ }_{ij}\sum_{\sigma=\uparrow,\downarrow}
  c^{\dag}_{i\sigma}c^{\   }_{j\sigma}
+\Delta_{ij}\, c^{\dag}_{i\uparrow}c^{\dag}_{j\downarrow}
+{\rm H.c.}
\right),
\label{eq: def cal H}
\\
&&
{\cal V}  =
\sum_{i}\sum_{\sigma=\uparrow,\downarrow}
\,V^{\ }_i\,c^{\dag}_{i\sigma}c^{\   }_{i\sigma},
\nonumber
\end{eqnarray}
that describes electrons hopping on a square lattice 
made of ${\cal N}$ sites with lattice spacing $a$.
The real valued hopping matrix $t_{ij}$ contains in its diagonal
the chemical potential $\mu_0$.
Its off diagonal elements describe nearest neighbor hopping ($t$),
next-nearest hopping ($t'$) etc., and yield the 
normal state dispersion $\varepsilon_{\bf k}$ in reciprocal space. 
The order parameter
$
\Delta_{ij}=
(\Delta_0/2)
\left(
 \delta_{i,j\pm\hat{\bf x}}
-\delta_{i,j\pm\hat{\bf y}}
\right)
$
enforces the $d_{x^2-y^2}$ symmetry.
The disorder potential $V_i$ describes $N$ uncorrelated $s$-wave scatterers
with the individual potential strength $V_0$. 
Moreover, $N$ is randomly distributed according to a Poisson distribution. 
Hence,
\begin{mathletters}
\begin{equation}
{\overline{V^p_i\,V^q_j}}^{\ }_c=
\rho\, V^{p+q}_0\, \delta^{\ }_{ij},
\qquad \rho={N\over{\cal N}},
\label{eq: def disorder}
\end{equation}
where the overline denotes disorder averaging (the subscript $c$
refers to the cumulant). 
Instead of parametrizing the disorder by $(\rho,V_0)$ 
it is more convenient to introduce the parametrization
$(\rho,g)$ where 
$
g
= \rho\, V^2_0.
$
We can then identify two distinct limits:
\begin{eqnarray}
&&
\rho\to\infty,
\qquad
g \hbox{ finite}.
\qquad \hbox{GL}
\label{eq: gauss limit on dis}
\\
&&
g\to g_{\rm r},\,
\qquad
\rho \hbox{ finite}.
\qquad \hbox{UL}
\label{eq: unitary limit on dis}
\end{eqnarray}
\end{mathletters}
The resonant disorder strength $g_{\rm r}=\infty$
when the normal state dispersion is particle-hole symmetric,
whereas it is a finite energy scale that depends on $\mu_0$, $t'$, etc.,
otherwise. To simplify notation we will assume
a particle-hole symmetric normal state dispersion from now on:
$g_{\rm r}=\infty$.

Effective field theories in Anderson localization are constructed to
compute the disorder average over products of the Green functions
$G(z)=({iz}-{\cal H})^{-1}$ at different Matsubara energies $z$.
For the problem at hand, the (bosonic) replicated canonical partition function
\begin{eqnarray}
&&
\Xi^{\ }_{2N,n}\propto
{1\over{\cal N}^N}
\sum_{i_1=1}^{\cal N}\ldots\sum_{i_N=1}^{\cal N}
\int{\cal D}[{\bm \phi}]\,
e^{
-{1\over2}
{\bm \phi}^{\rm T}
\left(
 {iz}
-{\bm {\cal H}}
\right)
{\bm \phi}
}
\nonumber\\
&&
\hphantom{\Xi^{\ }_{2N,n}}
=
\int{\cal D}[{\bm \phi}]\,
e^{
-{1\over2}
{\bm \phi}^{\rm T}
\left(
 {iz}
-{\bm {\cal H}}_0
\right)
{\bm \phi}
}
\left[{\cal A}(\rho,g)\right]^N,
\label{eq:def Xi}\\
&&
{\cal A}(\rho,g)=
{1\over{\cal N}}\sum_{i=1}^{\cal N}
e^{
-{1\over2}\sqrt{g\over\rho}\, 
\sum\limits_{a=1}^n\sum\limits_{\tau,\tau'=1}^2
\phi^{\ }_{ia\tau}\,(-)^{\tau}\delta^{\ }_{\tau,\tau'}\phi^{\ }_{ia\tau'}
},
\nonumber
\end{eqnarray}
yields, upon insertion of a source term,
the disorder average of the single-particle Green function $G(z)$ 
whereby disorder averaging is defined by weighing equally all
the spatial configurations that $N$ impurities can take on the lattice.
Here, ${\bm {\cal H}}={\bm{\cal H}}_0+{\bm{\cal V}}$ 
is a $2{\cal N}n\times2{\cal N}n$ matrix 
representation of Eq.~(\ref{eq: def cal H}) 
(in the Nambu representation with the Pauli matrices 
$\gamma_{1,2,3}$ to represent the particle-hole grading) 
and ${\bm \phi}$ is a 
replicated bosonic real field with $2\times{\cal N}\times n$
components. In order to make a connection between the GL and the UL
we need to relax the condition that $N$ is fixed
in Eq.~(\ref{eq:def Xi}). This is done by defining 
\begin{eqnarray}
&&
{\cal Z}^{\ }_{\alpha,n}=
\sum_{N=0}^\infty\, \Xi^{\ }_{2N,n}\, {\alpha^N\over N!}\, e^{-\alpha}
=
\int{\cal D}[{\bm \phi}]\,
e^{-S_0-S_1-\rho\,{\cal N}},
\nonumber\\
&&
S_0=
{1\over2}
{\bm \phi}^{\rm T}\left(iz-{\bm{\cal H}}^{\ }_0\right){\bm\phi},
\label{eq: Cal Z}\\
&&
S_1=
-\rho\,\sum_{i=1}^{\cal N}
\exp
\left[
-{1\over2}
\sqrt{g\over\rho}
\sum_{a=1}^n\sum_{\tau=1}^2
\phi^{\ }_{i\tau a}
(-)^\tau\delta^{\ }_{\tau\tau'}
\phi^{\ }_{i\tau'a}
\right].
\nonumber
\end{eqnarray}
This Poisson grand canonical partition function is normalized to one in
the replica limit $n\to0$. Denoting averaging with ${\cal Z}_{\alpha,n}$
by $\langle (\cdots)\rangle$, we deduce that the average
number of impurities is $\langle N\rangle =\alpha$ and that the 
standard deviation in the number of particles scales like 
$\sqrt{\alpha}$. From now on we will identify the impurity concentration
$\rho$ with $\alpha/{\cal N}$.
The interacting field theory described by Eq.~(\ref{eq: Cal Z})
has already been introduced by M\'ezard, Parisi and Zee 
as an effective field theory for random Euclidean matrices
\cite{Mezard99} (see also Refs.~\onlinecite{Friedberg75,Affleck83,Zee00}). 
Our derivation shows that
the interacting field theory with a Gaussian interaction appears very
naturally in the problem of Anderson localization with Poisson distributed
impurities.

According to Eq.~(\ref{eq: gauss limit on dis}), 
the GL for a white noise distribution of 
impurities  should be recovered as $\rho\to\infty$.
It is natural to expand the exponential of $S_1$ in this limit.
The first non-vanishing contribution to the action in the
grand canonical partition function is of order $g$
\begin{eqnarray}
&&
{\cal Z}^{\rm g}_{\alpha,n}=
\int{\cal D}[{\bm \phi}]\,
\exp
\left(
-S^{\rm g}_0
-S^{\rm g}_1
\right),
\nonumber\\
&&
S^{\rm g}_0=
{1\over2}
{\bm \phi}^{\rm T}\left(iz-{\bm{\cal H}}^{\rm g}_0\right){\bm\phi},
\qquad
S^{\rm g}_1=
-{g\over8}
\left({\bm\phi}^{\rm T}\gamma^{\ }_3{\bm\phi}\right)^2,
\label{eq: Cal Z Gauss}
\end{eqnarray}
aside from a shift in the chemical potential $\delta\mu=\sqrt{\rho\, g}$.
Equation (\ref{eq: Cal Z Gauss}) describes the effective 
action needed to calculate the average Green function of
the random BdG Hamiltonian ${\bm{\cal H}}^{\rm g}_0+{\bm v}$.
The superscript ``g'' indicates that the chemical potential
$\mu^{\rm g}_0=\mu^{\ }_0-\sqrt{\rho\, g}$ in ${\bm{\cal H}}^{\rm g}_0$.
The random potential ${\bm v}$ is white noise correlated 
and Gaussian distributed with vanishing mean and variance $g/4$, respectively.
The usual justification for neglecting terms of order
$\rho(g/\rho)^{3/2}$ is that they are irrelevant at the fixed point to
which $S^{\rm g}_0+S^{\rm g}_1$ flows. This fixed point is itself
described by a NLSM on the manifold Sp($2n$) in the limit of weak disorder,
i.e., when the mean free path in the Born approximation
$
\sim2t/g
$,
is much larger than the lattice spacing $a$ and for energies close to
the band center $\mu^{\rm g}_0=0$\cite{Senthil98,Simons00}.
At this fixed point the DOS
$\nu(\varepsilon)$ 
is only sensitive to the existence of nodes for not too small energies
\cite{Gorkov85,Fradkin86}. The physics of weak localization comes into play
below the energy scale 
$
\varepsilon^{\rm g}_1\sim
(B/A)\,\exp(-\pi\,B^2/2g)
$
at which a pseudo gap opens up.
Below an even smaller energy scale
$
\varepsilon^{\rm g}_2\sim
\varepsilon^{\rm g}_1\,[g/(A^2\,B^2)]\,\exp(-A^2)
$
random matrix theory (RMT) predicts that
the DOS becomes linear and insensitive to the nodes of $\Delta_{\bf k}$
(the prediction of RMT can break down for correlated disorder in which
 case sensitivity of the DOS to the nodes of $\Delta_{\bf k}$ 
 is recovered \cite{Nersesyan94,Mudry96,Huckestein00}).
Here, 
$B\propto\sqrt{2t\,\Delta_0}$ is the band width and 
$A\propto\sqrt{2t/\Delta_0}$ is the Dirac cone anisotropy.
For notational simplicity $2t=\Delta_0=1$ from now on.


According to Eq.~(\ref{eq: unitary limit on dis})
the natural expansion 
for small impurity concentrations is the virial expansion , i.e.,
one in powers of $\rho$ of
$\exp(-S_1)$ in Eq.~(\ref{eq: Cal Z}).
We have calculated the increment $\delta\nu^{(p)}(\varepsilon)$ 
in the DOS due to $p=1$ and $p=2$ impurities, respectively. 
As it should be we reproduce 
the known increment $\delta\nu^{(1)}(\varepsilon)$ 
induced by one impurity off and at resonance ($g=g_{\rm r}$) to first order in 
$\rho$\cite{Pethick86,Nagaosa95,Balatsky95,Khaliullin97}.
Off resonance the DOS crosses over to a linearly vanishing DOS upon
approaching $\varepsilon_{\rm F}$. {\it This is not so at resonance}.
We infer from the calculation of the averaged single
particle Green function $\langle G(z)\rangle$ up to second order in $\rho$
that the virial expansion is an expansion in the small parameter
$\alpha^{\rm u}\equiv\pi\rho/[z^2\ln(z^2)]$. 
Hence, the virial
expansion is not uniform in the Matsubara energy $z$.
This non-uniformity of the virial expansion is the counterpart
to the fact that in the GL the expansion in $1/k_{\rm F}\ell$ of
$\langle G(z)\rangle$ is really an expansion in the small parameter 
$\alpha^{\rm g}\equiv g\ln(z^2)/\pi$\cite{Nersesyan94}.
[The simplest guess, as suggested by the first terms in the virial expansion,
for a small expansion parameter that interpolates between these two limits is
$
\alpha^{\rm eff}\equiv 
\alpha^{\rm u}\alpha^{\rm g}/
(\alpha^{\rm u}+\alpha^{\rm g})
$.]
The non-uniformity of the perturbative expansions in $\alpha^{\rm g,u}$ 
reflects the importance of quantum interferences already at the
level of the single-particle Green function.
Both perturbative breakdowns below the thresholds $\varepsilon^{\rm g,u}_1$
are signatures of the existence
of isolated nodes of $\Delta_{\bf k}$ in the 2d Brillouin zone
that require non-perturbative approaches to access the DOS arbitrarily close
to $\varepsilon_{\rm F}$.
We have not been able to find a counterpart to the Sp($2n$)
NLSM that describes the GL and is amenable to a calculation
of the DOS arbitrarily close to $\varepsilon_{\rm F}$. 
On the other hand, comparison of the expansions in $\alpha^{\rm g}$
and $\alpha^{\rm u}$ up to second order suggests that the GL and UL are 
related by $g/\pi\leftrightarrow\pi\rho$ 
and $\ln(z^2)\leftrightarrow1/z^2\ln(z^2)$.
If this relationship holds to all orders 
and extends to the non-perturbative regime as well it would allow 
access to 
$\lim_{\varepsilon\to\varepsilon_{\rm F}}\nu(\varepsilon)$
in the UL from knowledge of 
$\lim_{\varepsilon\to\varepsilon_{\rm F}}\nu(\varepsilon)$
in the GL. We will argue from a detailed study of a variational approach that
this duality relating the GL and UL indeed holds.

For practical purposes $\varepsilon^{\rm g,u}_1$ can be considered small
and the crossover of the DOS between the GL and UL is 
needed when $\varepsilon>{\rm max}\,\varepsilon^{\rm g,u}_1$.
This can be achieved by treating the interacting theory (\ref{eq: Cal Z})
within a variational approximation. 
We thus replace the kernel of
$S_0+S_1+\rho\,{\cal N}$ in Eq.~(\ref{eq: Cal Z})
by the non-interacting, translationally invariant, and replica diagonal
kernel ${iz}'-{\cal H}^{\rm sc}$, $z'\equiv z+\bar z$. 
In reciprocal space 
$
{\cal H}^{\rm mf}_{\bf k}=
 \varepsilon'_{\bf k}\, \gamma^{\ }_3
+\Delta'_{\bf k}\,\gamma^{\ }_1
$,
$
\varepsilon'_{\bf k}\equiv
 \varepsilon^{\ }_{\bf k}
-\mu_0
+\bar\varepsilon,
$
and
$
\Delta'_{\bf k}\equiv
 \Delta^{\ }_{\bf k}
+\bar\Delta.
$
The self-consistent parameters $\bar z$, $\bar\varepsilon$, and
$\bar\Delta$, i.e., the disorder induced self-consistent shifts in the 
self-energy, chemical potential, and gap, respectively, 
are obtained by solving the self-consistent equations
\begin{mathletters}
\begin{eqnarray}
&&
\Omega({z}')=
\rho\,
{\bar z\over \bar z^2+\bar\varepsilon^2+\bar\Delta^2},
\label{eq:SC Omega'}
\\
&&
{\cal E}({z}')=
\rho\,
{\bar\varepsilon\over \bar z^2+\bar\varepsilon^2+\bar\Delta^2}
-\sqrt{\rho\over g},
\label{eq:SC cal E'}  
\\
&&
\Delta({z}')=
\rho\,
{\bar\Delta\over \bar z^2+\bar\varepsilon^2+\bar\Delta^2},
\label{eq:SC Delta'}
\\
&&
\left(
\Omega  ,
{\cal E},
\Delta  
\right)
({z}')
=
{1\over{\cal N}}
\sum_{\bf k}
{
\left(z',\varepsilon'_{\bf k},\Delta'_{\bf k}\right)
\over
 (z')^2
+(\varepsilon'_{\bf k})^2
+(\Delta'_{\bf k})^2
}.
\label{eq: O' E' D'}
\end{eqnarray}
\end{mathletters}
The self-consistent DOS is
$
\nu^{\ }_{\rm sc}(\varepsilon)=
$
$
\lim_{\eta\to0}
$
$
\lim_{{iz}\to\varepsilon+i\eta} 
$
$
{\rm Im}\left[i\,{\rm sgn}(\eta)\,\Omega(z')\right].
$
It is sufficient to consider 
Eqs.~(\ref{eq:SC Omega'},\ref{eq:SC cal E'})
alone since $\Delta_{k_x,k_y}=-\Delta_{k_y,k_x}$
allows $\bar\Delta=0$. 

That details of disorder matter in 2d $d$-wave superconductors
(see Ref.~\cite{Atkinson00})
already follows from the variational approach. Indeed,
Eqs.~(\ref{eq:SC Omega'},\ref{eq:SC cal E'}) are not so much sensitive to
the energy $z$ as they are to the chemical potential $\mu_0$
when it comes to describing the crossover from the UL to the GL.
Equations (\ref{eq:SC Omega'},\ref{eq:SC cal E'})
reduce to the two well-known self-consistent equations 
\begin{mathletters}
\label{eq: sc UL+GL}
\begin{eqnarray}
&&
\rho\to\infty,
\quad
g<\infty,
\quad
\mu_0=\sqrt{\rho\, g}
\Rightarrow
\Omega(z')={\bar z\over g},
\quad
\bar\epsilon=\mu_0,
\nonumber\\
&&
\label{eq: sc GL}
\\
&&
g\to \infty,
\quad
\rho<\infty,
\quad
\mu_0=0
\hphantom{\rho\, g\ }
\Rightarrow
\Omega(z')={\rho\over\bar z},
\quad
\bar\epsilon=0,
\nonumber\\
&&
\label{eq: sc UL}
\end{eqnarray}
\end{mathletters}
in both the 
GL\cite{Gorkov85,Fradkin86}
and the 
UL\cite{Schmitt86,Hirschfeld88}.
It is thus imperative to scale $\mu_0$ with $\sqrt{\rho\, g}$
for $\bar\varepsilon$ to be tuned away from the UL resonance
$|\bar\varepsilon|\ll|\bar z|$.

The self-consistent Eq.~(\ref{eq: sc GL})
does not capture the interplay between the physics of
localization and that of level repulsion below the energy scale
$\varepsilon^{\rm g}_1$\cite{Nersesyan94,Senthil98}.
The same should be true
of Eq.~(\ref{eq: sc UL})
below the energy scale $\varepsilon^{\rm u}_1$ at which the virial
expansion breaks down. We are going to argue that the physics of 
localization and level repulsion results in a diverging DOS upon
approaching $\varepsilon_{\rm F}$ in the UL. Our argument relies first
on identifying a duality relating the perturbative expansions
of the self-energy in the GL and UL valid at {\it large} energies
relative to the disorder strength. We then postulate that the same
duality extends to {\it small} energies relative to the disorder strength.
Let $z$ be a large Matsubara energy, i.e., $g$ ($\rho$)
is sufficiently small so that 
level repulsion can be ignored
and the physics of localization is that of the standard orthogonal symmetry
class. In this regime, we can safely remove the self-consistency condition in
\begin{mathletters}
\begin{eqnarray}
\Omega(z+\bar z)=&&
  \sum_{p=0}^\infty 
  \cases{
  {g^p   \over p!}\,\left[\Omega(z')\right]^{+p}\,
  (\partial^p_z\Omega)(z),
  &  GL,\cr
  {\rho^p\over p!}\,\left[\Omega(z')\right]^{-p}\,
  (\partial^p_z\Omega)(z),
  &  UL,\cr
  }
\label{eq:series}
\end{eqnarray}
by substituting 
$\Omega(z+g\Omega(z+\dots)\cdots)$ 
[$\Omega(z+\rho/\Omega(z+\dots)\cdots)$]
with
$\Omega(z)+(\partial_z\Omega)(z)\Omega(z)\,g+{\cal O}(g^2)$ 
[$\Omega(z)+(\partial_z\ln\Omega)(z)\,\rho+{\cal O}(\rho^2)$].
In this way, the self-energy reduces to the perturbative expansion
\begin{eqnarray}
\Omega(z+\bar z)\approx&&
  {z\ln(z^{-2})\over\pi}\!
  +{2z[\ln(z^2)+2]\over\pi}
   \cases{
   {\ln(z^2)\over2}\,{g\over\pi}& \cr
   {\pi\,\rho\over2z^2\ln(z^2)} &\cr
   }
\nonumber\\
-&&{2z\over\pi}
  \sum_{p=2}^\infty 
  \cases{
   \beta^{\rm g}_p\,
  [\ln(z^2)]^{+p}\,
  \left(g\over\pi\right)^{p},
  &  GL.\cr
   \beta^{\rm u}_p\,
  [z^2\ln(z^2)]^{-p}\,
  \left(\pi\rho\right)^{p},
  &  UL.\cr
  }
\label{eq:series'}
\end{eqnarray}
\end{mathletters}
Here, the combinatoric coefficient $\beta^{\rm g,u}_p$ depends
on the order in $g,\rho$ to which the self-consistency has been
truncated. For example, the substitution $\Omega(z')=\Omega(z)$ gives
$\beta^{\rm g}_p=1/p(p-1)=\beta^{\rm u}_p$ 
and accounts only for the pure
rainbow diagrams in the expansion of the Dyson equation for the
self-energy in the GL. Truncation to higher order in $g$ accounts for 
diagrams that are not purely of the rainbow type but in which impurity lines 
never cross. Neglecting diagrams with crossing of
impurity lines is only justified for sufficiently weak disorder at a fixed
energy\cite{Nersesyan94}. 
Correspondingly, the condition on the disorder strength for the physics
of localization and level repulsion to be unimportant is
\begin{mathletters}
\label{eq: radius conv rainbow}
\begin{eqnarray}
&&
-\alpha^{\rm g}\equiv
\frac{g}{\pi}
\ln (1/z^2)<1
\Longleftrightarrow
|z|>|\bar z^{\rm g}_0|,
\quad {\rm GL}.
\label{eq: z_1 GL}
\\
&&
-\alpha^{\rm u}\equiv
\quad
\frac{\pi\,\rho}{z^2\ln(1/z^2)}<1
\Longleftrightarrow
|z|>|\bar z^{\rm u}_0|,
\quad {\rm UL}.
\label{eq: z_1 UL}
\end{eqnarray}
\end{mathletters}
Note that the expansions in the GL and UL are related
\begin{equation}
\frac{g}{\pi} \leftrightarrow \pi\,\rho,
\qquad
\ln(z^2)  \leftrightarrow \frac{1}{z^2\ln(z^2)}.
\end{equation}
By undressing the propagator in Eq.~(\ref{eq:series'}),
two expansions of the same scaling form
$
\Omega(z')-\Omega(z)=
2(z/\pi)\{{1\over2}[\ln(z^2)+2]\alpha^{\rm g,u}-f(\alpha^{\rm g,u})\}
$ 
have been derived for {\it large} energies in the GL and UL, respectively.
We now assume that the self-energy can also be written
as some scaling function $\tilde f$
that can be expanded in powers of $\alpha^{\rm g,u}$ outside
of the radius of convergences (\ref{eq: radius conv rainbow}), 
i.e., at {\it low} energies.
The scaling function $\tilde f$ should be specified by the 
symmetry class CI, i.e., the physics of level repulsion induced by
the particle-hole symmetry of a time-reversal and spin-rotation symmetric
superconductor. This scaling function is  not completely unknown since
RMT predicts that
$\tilde f(\alpha^{\rm g})\propto\alpha^{\rm g}$ for $|\alpha^{\rm g}|\gg1$.
[$(-2z/\pi)(g \ln z^2/\pi)$ is the combination that, 
upon analytical continuation, leads to
$\nu^{\rm g}(\varepsilon)\propto |\varepsilon|$ 
arbitrarily close to $\varepsilon=0$.]
By the duality assumption
$\tilde f(\alpha^{\rm u})\propto\alpha^{\rm u}$ must also hold
outside of the radius of convergences (\ref{eq: z_1 UL}) 
[leading to the combination $(-2z/\pi)(\frac{\pi\rho}{z^2\ln z^2})$].
Upon  analytical continuation, duality then predicts a diverging DOS
$\nu^{\rm u}(\varepsilon)\propto1/|\varepsilon|\ln^2|\varepsilon|$
arbitrarily close to the Fermi energy $\varepsilon^{\ }_{\rm F}$ in the UL.
The robustness of this conjecture requires a careful study of the role
played by diagrams with crossing impurity lines (the Goldstone modes in NLSM
terminology) in the UL that is beyond the scope of this paper.

Our main conclusion is that the crossover between the resonant UL and the GL
in a 2d dirty $d$-wave superconductor is not controlled by energy only
but by the chemical potential and the strength of
the disorder, which together tune out of the resonance.
A variational approach yields self-consistent equations that interpolate 
between the GL and UL. In practice these self-consistent equations describe
accurately the DOS since the energy scales
signalling breakdown of perturbative expansions are expected to be rather 
small. We have established under what conditions the diverging DOS
$1/|\varepsilon|\ln^2|\varepsilon|$ predicted by P\'epin and Lee 
in Ref.~\onlinecite{Pepin98} can arise. A duality between the UL and the GL
that holds in the regime of applicability of the variational approach 
must extend to the non-perturbative regime. In this way the apparent 
dichotomy between a DOS controlled solely by RMT in the GL and one
controlled solely by the nodes of the gap is elegantly resolved.
Only the leading functional dependence on energy is predicted to be universal
by RMT. Accordingly, we cannot predict the prefactor of the diverging DOS 
whereas P\'epin and Lee state that this prefactor is exactly given 
by $\rho$, coincidentaly the prefactor for independent impurities.

We thank Patrick Lee and Catherine P\'epin for valuable discussions. 
This work was supported by the NSF under grant No.~DMR-98-76208,
and the Alfred P.~Sloan Foundation (CC) and by an ESF/FERLIN grant (CM).

\narrowtext

\end{document}